# Practical Context Awareness: Measuring and Utilizing the Context Dependency of Mobile Usage

Ahmad Rahmati[1], Clayton Shepard[1], Chad Tossell[2], Lin Zhong[1], Philip Kortum[2]
[1] Department of Electrical & Computer Engineering, [2] Department of Psychology, Rice University

## Abstract

Context information brings new opportunities for efficient and effective applications and services on mobile devices. A wide range of research has exploited context dependency, i.e. the relations between context(s) and the outcome, to achieve significant, quantified, performance gains for a variety of applications. These works often have to deal with the challenges of multiple sources of context that can lead to a sparse training data set, and the challenge of energy hungry context sensors. Often, they address these challenges in an application specific and ad-hoc manner. We liberate mobile application designers and researchers from these burdens by providing a methodical approach to these challenges. In particular, we 1) define and measure the context-dependency of three fundamental types of mobile usage in an application agnostic yet practical manner, which can provide clear insight into the performance of potential application. 2) Address the challenge of data sparseness when dealing with multiple and different sources of context in a systematic manner. 3) Present SmartContext to address the energy challenge by automatically selecting among context sources while ensuring the minimum accuracy for each estimation event is met.

Our analysis and findings are based on usage and context traces collected in real-life settings from 24 iPhone users over a period of one year. We present findings regarding the context dependency of the three principal types of mobile usage; visited websites, phone calls, and app usage. Yet, our methodology and the lessons we learn can be readily extended to other context-dependent mobile usage and system resources as well. Our findings guide the development of context aware systems, and highlight the challenges and expectations regarding the context dependency of mobile usage.

## 1. Introduction

Modern mobile systems such as smartphones and tablets are already important part of our lives. They are not only computationally powerful but also have a rich capability to sense their external and internal environment. Similar to the definition by Schilit et al. in [1], we refer to the last known condition of these environments collectively as *context*. Context dependency can be broadly defined as a set of strict or probabilistic rules and relations between context(s) and the outcome [2].

Context has in the past been widely exploited to provide more usable mobile applications and services, such as content adaptation [3, 4], user interaction [5], and information delivery [6, 7]. Context has also been widely exploited to provide enhanced system efficiency and performance, such as for energy management [8, 9] and network selection [10]. These designs exploit the context dependency of mobile usage and/or mobile resources for specific purposes, and show significant, quantified, performance gains.

Context aware systems often have to deal with two fundamental challenges. First: dealing with multiple sources of context is challenging; due to the curse of dimensionality [11], simply treating them as a multidimensional vector results in a sparse training set. Second: liberal application of context can quickly drain the devices battery, as some context sensors are extremely energy hungry. To address the sparseness challenge, existing work often limit the number of context sources, e.g. to one [8] or two [9], and/or employ ad-hoc or expert solutions to combine multiple sources of context, e.g. [10]. To address the energy challenge, they often employ ad-hoc schemes along one or more of these lines: reducing the frequency of accessing costly context [12-14], avoiding them altogether [10, 12, 13], or substituting them with other context [15-18].

Ad-hoc and application specific approaches towards these challenges mean that the designers need to design and evaluate a new solution for every context-based system. Furthermore, before designing and evaluating their application or service, its designers can only guess its performance outcome. Our work is liberating in this regard. We provide a methodological solution for using multiple and various sources of context, while managing their energy costs. We provide a formal yet practical definition of context dependency, which provides insight into the performance of applications while remaining application agnostic. We measure the context dependency of three principal types of mobile usage, using unprecedented real-life context and usage traces collected from 24 iPhone users over one year. The mobile usage we focus on are visited websites, phone calls, and app usage[1]. We utilize context information from sensors built into the phone (i.e. real-time clock, Cell ID, Accelerometer, and GPS), as well as the phone's last known usage state (i.e. application, web, and phone use). Yet, our methodology and the lessons learned can be extended to other context and usage as well. In particular, we make four

---

[1] Note that we use the words app and application differently. App refers to applications that are installed on the phones, either built-in or obtained from the App Store. Application refers to its more general meaning, i.e. use case. Similarly, application agnostic means not dedicated to a single service or purpose.

**Table 1. Data samples collected from the 24 users, during one year of logging**

| Type of usage | Total samples | Mean samples per user |
|---|---|---|
| Websites visited | 17,000 | 700 |
| Phone calls | 54,000 | 2,300 |
| Applications launched | 508,000 | 21,200 |

major contributions towards quantifying and measuring the context dependency of mobile usage:

First, in Section 3, we identify *estimation accuracy* based on maximum a posteriori (MAP) estimation as an application agnostic yet practical, measure for context dependency, In contrast with other theoretical metrics that are applicable applied to multinomial data, such as entropy as a measure of uncertainty and pseudo R square as a measure of correlation, estimation accuracy provides practical insight into the performance of many potential applications, while remaining application agnostic. To allow the efficient calculation of posterior probabilities, we present and compare the performance of several forms categorizing and binning of context measurements into a limited number of categories, for both continuous and discrete context sources. Furthermore, we address the challenge of data sparseness when dealing with multiple sources of context by comparing classifier combination methods.

Second, in Section 4, we present a series of interesting findings regarding the context dependency of mobile usage, as follows: 1) The effectiveness of different context varies based on the usage to be estimated, as well as the number of accepted responses. Yet, combining multiple sources of context uncovers their combined strength. 2) We find that even though multiple context sources are dependent, Bayesian Combination performs well for combining context information. 3) The context-dependency of usage remains relatively constant even for durations of one to three months, instead of the full 12 months. This indicates that a smaller data set would be sufficient for context-awareness. 4) Supervised Binning can greatly increase estimation accuracy by keeping a large number of samples in each category or bin, while allowing fine molding of the bins. 5) Even though users are diverse in their usage, we are able to show substantial context dependency among all of them.

Third, in Section 5, we present SmartContext, a framework to dynamically or statically optimize the cost / accuracy tradeoffs of context awareness, while ensuring a minimum accuracy for each estimation event. SmartContext takes advantage of the classifier combination algorithms we have explored that have little overhead. We show that by utilizing energy hungry context only at uncertain times, SmartContext can achieve an estimation accuracy within 1% of the maximum possible accuracy, while significantly reducing energy costs by 60% or more.

Fourth, in Section 6, we present and evaluate several sample applications that benefit from context dependency of mobile usage. These applications highlight the practical value of estimation accuracy as a measure of context dependency, and attest to the effectiveness of context for estimating usage. Our best performing methods, i.e. using Supervised Binning and Bayesian combination, consistently outperform common non-context-based methods.

## 2. Data Collection

Studying context dependency can be extremely challenging, as it needs a large trace collected in real life user studies. In this section, we describe the methodology used to collect and analyze the usage and context data from 24 iPhone users. We have already presented the details of the data collection in [19]. In this section, we provide information relevant to this study.

### 2.1 Field Study Participants

The 24 participants were studied continuously for one year, from February 2010 to February 2011. All of them were undergraduate students at a small private university, located in a major metropolitan area of the USA. In general, they were representative of college students in terms of age (average age: 19.7, deviation: 1.1) and gender. They lived on campus and had a PC or laptop at their residence, in addition to access to the university's computing labs.

As compensation, each participant received a free iPhone as well as free service throughout the duration of the study, including 450 voice call minutes per month, unlimited data, and unlimited SMS. We helped all participants port their phone numbers to and they were required to use the outfitted iPhones as their primary phone. They were not given specific instructions on how to use the device, other than to use it as they would normally use their phones.

### 2.2 Logger Design and Implementation

While extensive logging of PC usage has been reported in past literature, privacy concerns and battery lifetime limitations, have limited the scope of mobile phone based studies. Indeed, privacy concerns and/or significantly reduced battery lifetime is likely to impact usage, thus the usage data would not accurately real life user behavior [20, 21]. Our study mitigates these concerns by limiting energy consumption and addressing user privacy concerns through one way hashing and on device data processing, as well as by *partitioning*, i.e. dividing the research team so that the data analysis and logger development team do not know or directly interact with the participants, in order to avoid linking data to the actual users. The key component of the study is an in-device, programmable logging software that collects iPhone usage and context *in situ*. To run the iPhone logger continuously in the background, we had to jailbreak the iPhone. The main logger daemon is written as a bash shell script and utilizes components written in various languages, including C, Perl, awk, SQL, and objective C, altogether comprising ~2000 lines of code.

The logger records a plethora of context information. For this work we focus on logs regarding usage and context. The visited websites, app used, and phone calls are record-ed by the phone's operating system, and our logger piggy-backs on the phone's logs by periodically recording them. Further, whenever the phone's CPU is not asleep, at 15 minute intervals, the logger records the GPS location, cell ID, and a 15 second recording from the accelerometer at 25hz. The GPS location data is collected using Apple's framework, which reports the GPS location if available, and, if not, the estimated location based on visible cell towers and strengths. The logger attempts to retrieve the location until the reported accuracy is less than 100m, or the location has been updated (by the framework) 3 times, in order to avoid draining the battery, yet still retrieve accurate location data. For cell ID location, we query the phone's GSM modem using the AT command set, returning the currently associated cell ID.

The collected data is recorded on the phones, and transferred nightly to our servers on a secure connection. Our logger has recorded thousands of usage samples through the study, as presented in Table 1. Due to the extremely large size of the traces, it is often necessary to process them sequentially. Therefore, we developed most of the tools to process them using the Perl language and Bash scripts. We also took advantage of several open source tools for this purpose, including Cluster 3.0 [22, 23].

### 2.3 Collected Data

#### 2.3.1 Usage

We look into the three fundamental types of mobile usage: phone calls, web usage, and app usage. We limit the number of usage categories considered to 100, similar to what we did with discrete context. This simplifies data processing, and can even increase accuracy by reducing usage cases with too few samples. We chose the number 100 based on the CDF of usage, covering 87%, 93%, and 99% of web, phone, and app usage.

We consider web usage as the independent websites a user visits, as presented by their domain names. We consider each visited domain as one entry, irrespective of the number of web pages under that domain. This would include the top level domain (TLD) and the first hierarchical subdomain, e.g. www.example.com/url/ would be counted as example.com. For phone calls, while our logger does not record actual phone numbers, it records a one-way hash that uniquely identifies each phone number. We consider all phone calls including ones that have a length of zero, indicating no conversation. For application usage, we consider all applications the user utilizes, including built-in ones and those obtained from the App Store. We do not consider the home screen as an application, even though it is implemented as an application on the iPhone platform.

#### 2.3.2 Context

We considered several sources of context in two broad categories; *sensor context* that is sensed through the phone sensors, and *usage context* that is last known usage state of the phone. The sensor context we utilize are time&day, movement (accelerometer power), cell ID location, and GPS location. The usage contexts we utilize are the prior visited website, phone call, and application.

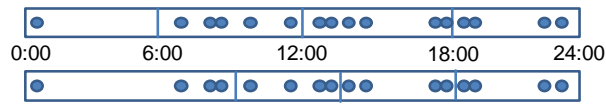

**Figure 1: Equal width discretization (top) would result in more natural boundaries. Equal frequency discretization (bottom) would use bins more efficiently and prevent too few samples in some bins.**

For time&day, we separate weekends and weekdays, but otherwise treat days as the same. Separating weekends from weekdays not only makes intuitive sense, but our testing indicated that it performed better than treating all days the same. Therefore, with a one-minute resolution, time&day is a continuous number between 0 and 2880, to account for a two day period (a weekday and a weekend).

For movement, we calculate the log of the power of the accelerometer readings. The reason we utilize the log of power, instead of absolute power, is the distribution of power readings that is close to the power law. More than 99% of the $log(p)$ entries fall between 0.1 and 10000, and the range is therefore limited accordingly.

For GPS location, we utilize the most accurate location provided by the iPhone API, which is provided in the geographic coordinate system, i.e. latitude and longitude. For cell ID location, we utilize the (single) cell ID reported by the phone.

## 3. Quantifying Context-Dependency

As previously mentioned, context dependency can be broadly defined as a set of strict or probabilistic rules and relations between two often discrete variables context(s) and outcome [2]. Multiple theoretical, application agnostic metrics exist for measuring the relationship of such variables. These include entropy as a measure of uncertainty, and Pseudo R Square as a measure of correlation [24]. Yet, neither entropy nor Pseudo R Square can provide practical insight into the performance of context-aware applications.

We present estimation accuracy, based on maximum a posteriori probability (MAP) estimation, as our measure of choice for context dependency. Estimation accuracy can provide practical insight into the performance of many potential context-aware applications, while remaining application agnostic. In this section, we provide practical methods to calculate the a posterior probability of an outcome ($g$) given context ($x$), or $P(g|x)$, from one or multiple continuous or discrete (multinomial) context sources.

### 3.1 Formal Definition

The use of context information can help increase the estimation accuracy of MAP estimation. MAP estimation works as follows. Assumes $g$ takes value from a finite set $\{g_1, g_2, …, g_k\}$. Knowing the posterior probability of every possible outcome, $g$, under contextual information $x=(x_1, x_2, …, x_n)$, the optimal estimation for the outcome, $\hat{g}$, is

$$\hat{g}(x) = \underset{i}{\operatorname{argmax}} P(g = i|x)$$

where $P(g|x)$ is the a posteriori probability. Now the expected estimation accuracy is $\sum_x P(g = \hat{g}(x)|x)P(x)$, which should be higher than $P(g = \hat{g})$, the expected estimation accuracy without the contextual observation $x$.

For many applications, the cost of a false negative is considerably higher than false positive. Therefore, providing multiple responses (i.e. best guesses) can be beneficial. Such responses would be in the form of $\hat{g} = \widehat{g_1} \cup \widehat{g_2} \cup \ldots$. For example, in application preloading, the system could preload multiple applications to reduce the chance of not having the next application preloaded. We can use the same definition to allow multiple responses. In this case, the expected estimation accuracy would be

$$\sum_x P(g \in \hat{g}(x)|x)P(x).$$

## 3.2 Calculating Posterior Probabilities

The key to MAP estimation is the accurate calculation of the a posteriori probability distribution $P(g|x)$.

It may initially appear straightforward to calculate $P(g|x)$; simply dividing the number of times each possible outcome $g_i$ has occurred under context conditions $x$ by the number of times $x$ has been observed in total. Recall that $g$ takes value from a finite set, $\{g_1, g_2, \ldots, g_k\}$. However, if the number of times $x$ has been observed is small, the estimates of $P(g|x)$ are unreliable [25]. Due to the large number of possible context combinations, and the possibility of having few, or no prior samples in a given context, posterior probability estimates may become inaccurate or impossible. This is true even for individual context sources, but can become significantly worse if multiple context sources are treated as multiple dimensions, due to the curse of dimensionality [11]. In this subsection, we present the methods we use to address this challenge, for individual and multiple context sources.

### 3.2.1 Individual Context

For individual context, we employ Laplace Correction [26, 27] to reduce the negative impact of too few observations under some context conditions. Instead of calculating

$$P(g_i|x) = \frac{count(g_i|x)}{count(x)}$$

we employ Laplace Correction and calculate $P(g_i|x)$ as

$$P(g_i|x) = \frac{count(g_i|x) + m.P(g_i)}{count(x) + m}$$

where $m$ is the number of possible outcomes of g.

Note that Laplace Correction only smoothes out the estimate of $P(g_i|x)$ when there are a small number of context samples. The effect of Laplace Correction is negligible for those if are significantly more samples than outcomes, i.e. $count(x)>>m$.

### 3.2.2 Multiple Context

As presented earlier, treating the context space as a multi dimensional space, with each dimension corresponding to one context source, will result in an unacceptably sparse data set due to the curse of dimensionality. In order to address this challenge for calculating posterior probability, $P(g|x)$, we employ classifier combination techniques. They enable us to treat each context source as a separate one-dimensional predictor, and combine multiple $P(\hat{g}_i|x_n)$ into $P(\hat{g}_i|x_1, x_2, \ldots, x_n)$. In other words,

$$P(\hat{g}_i|x_1, x_2, \ldots, x_n)$$
$$= combination(P(\hat{g}_i|x_1), P(\hat{g}_i|x_2), \ldots, P(\hat{g}_i|x_n))$$

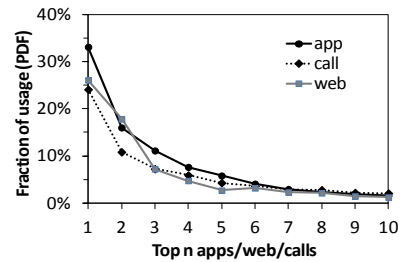

Figure 2: Estimating mobile usage is challenging due to the Power Law distribution of usage (average for all users); even given context based evidence, it is improbable for the posterior probability of a usage to rise above the more common usage.

We explore three prominent classifier combination techniques. The first is *Simple* or *Naïve Bayesian*, which works under the assumption that different sources of context are conditionally independent. The Bayesian rule calculates

$$P(g|x) = \frac{P(x|g)P(g)}{P(x)}$$

where $\quad P(x|g) = P(x_1,|g).\ldots.P(x_n|g)$

The assumption that different sources of context are independent does not necessarily hold. Even so, Simple Bayesian is known to often perform well even without this condition [28, 29]. We therefore evaluate the performance of Simple Bayes alongside other methods. Similar to individual context, we utilize Laplace Correction to reduce the ill effects of too few samples in calculating each $P(x,|g)$.

The second combination technique we explore is the *Maximum Rule*. The probability of each outcome is reported proportional to the maximum probability of that outcome among all classifiers, so that the sum of probabilities remains equal to one. Formally,

$$P(g_i|x_1, x_2, \ldots, x_n)$$
$$\propto max(P(g_i|x_1), P(g_i|x_2), \ldots, P(g_i|x_n))$$

For example, if one classifier selects outcome A with a 80% posterior probability, and two other classifiers select outcome B with 70% and 60% posterior probabilities, outcome A will be selected, and a posterior probability proportional to 80% is reported. The Maximum Rule is known to be highly sensitive to noise [30], when one classifier may be producing a high confidence due to noisy data or too few samples.

The third combination technique we explore is the *Mean Rule*, also known as the *Average Rule*. It calculates the probability of each outcome as the average of the reported probabilities by each of the classifying methods. Formally,

$$P(g_i|x_1, x_2, ..., x_n)$$
$$= mean(P(g_i|x_1), P(g_i|x_2), ..., P(g_i|x_n))$$

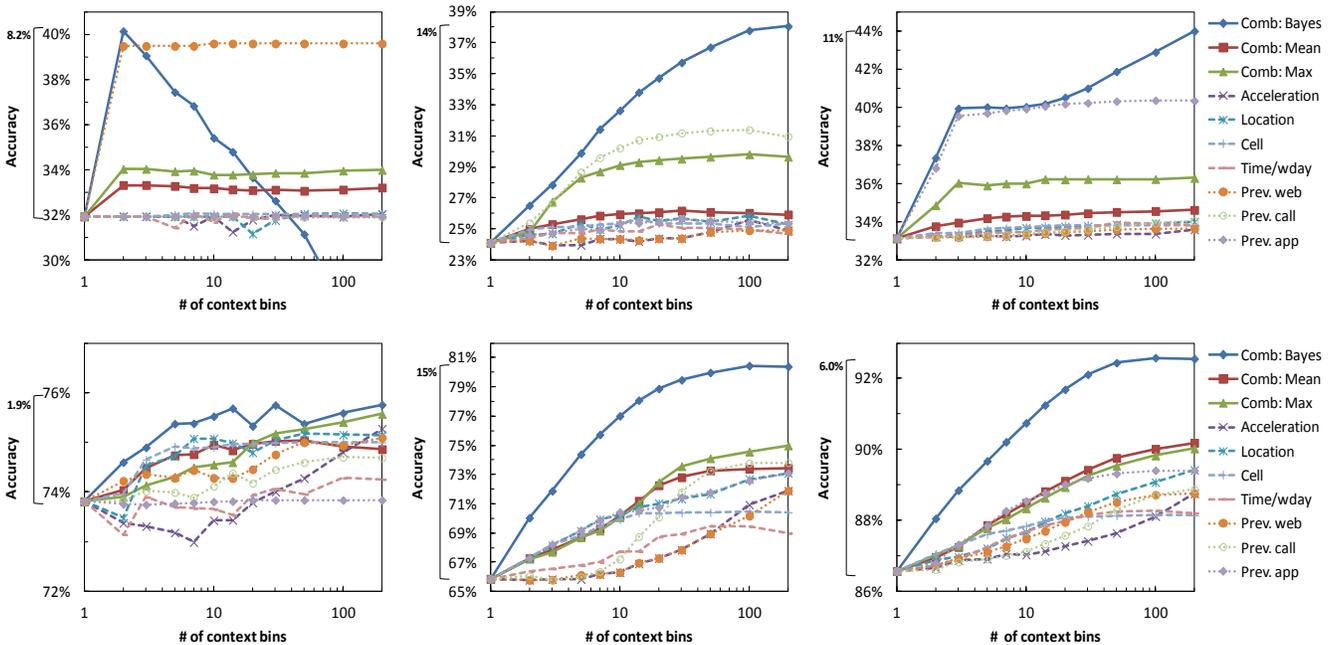

**Figure 3: Context dependency of web (left), phone (middle), and application (right) usage, presented as the accuracy of prediction, for 1 (top), and 10 (bottom) responses. One bin means no context information was used.**

The Mean Rule is especially resilient to noise [31], and most useful when the classifiers are highly correlated.

### 3.3 Discretizing Context

Based on our definition of context dependency, we need to discretize continuous sources of context. Continuous sources of context inherently have a single or multidimensional structure to them, where samples close to each other are related. Such a structure allows for the use of unsupervised clustering techniques to discretize them. There have been many methods in literature for unsupervised clustering of single and multi-dimensional entries, and consequently creating the discrete cases. Furthermore, there is often an option of adding an *equal frequency* constraint, as opposed to *equal width* discretization, as shown in Figure 1. An equal number of samples per cluster would guarantee efficient use of clusters, and prevent a too few number of samples in some clusters producing inaccurate results. On the other hand, the equal sample constraint may artificially limit the boundaries in the clustering algorithm, resulting in inefficient clusters. Both equal width and equal frequency discretization is straightforward for one dimensional context, such as time and movement (accelerometer power).

For continuous context in multiple dimensions, i.e. location in our case, we refer to clustering literature to find a suitable unsupervised clustering algorithm. The resulting clusters would in turn become the categories. We have two requirements for such an algorithm. We chose the popular k-mean clustering method because it satisfies our constraints; it clusters close samples together by minimizing the Within-Cluster Sum of Squares (WCSS), and it clusters entries into our desired number of clusters (k).

For equal frequency discretization of multidimensional context, we propose and evaluate an extension to k-means. Assuming $m$ entries and a cluster size of $n$ ($m=k.n$), our algorithm works as follows. First, a regular k-mean clustering is performed on the entries. The biggest cluster, i.e. the one with the most entries, is then selected, and the closest $n$ entries to the mean are assigned to that cluster. The remaining entries are clustered again, using the same method (i.e. $m–n$ entries into $k–1$ clusters). This is repeated until all entries are clustered. To retain meaningful results and prevent overlapping entries to be placed into different clusters, the equal size constraint is relaxed when two entries overlap or the resulting cluster radius is too small (< 5 meters).

### 3.4 Binning Context

Even for context that is already categorical, in order to efficiently calculate $P(g/x)$, it is necessary to limit the number of possible categories for $x$, i.e. by grouping together some of the categories. We use *binning* to refer to the process of reducing the number of context categories. We define Simple Binning as follows. For binning categorical context into $n$ bins, we simply choose the most popular $n-1$ categories, and group all other categories as the $n$'th bin. This is especially reasonable if the distribution of context follows the power law, which is often the case. For consistency, for continuous context, we define Simple Binning into $n$ bins as discretizing the context into $n$ categories.

The number of bins chosen for any context involves an inherent tradeoff; more bins can allow finer molding of bins and more accurate posterior probability calculations, but at the same time would result in fewer samples per category, increasing noise and reducing the accuracy of posterior probability calculations. Clearly, not only does the number of bins affect accuracy, but also how the bins are formed can affect accuracy. Supervised Binning can potentially make the best of both worlds in the above tradeoff by identifying and binning together categories that have similar outcomes. Therefore, supervised binning can allow finer molding of bins, increasing accuracy, without reducing the number of samples per bin that would increase noise and reduce accuracy. Supervised Binning can also be performed on continuous context, by first discretizing it into a larger number of categories.

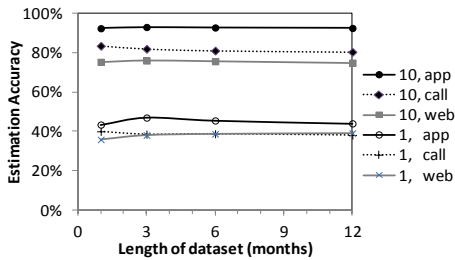

**Figure 4: Affect of seasonal variation of mobile usage on its context dependency is small, and one to three months of training logs is sufficient. Estimation accuracy for with one and ten responses, calculated on trace durations of one, three, six, and 12 months.**

There are two methods to perform Supervised Binning. First, one can either use the derived partitions from a classifier tree as the bins. An optimal classifier tree using (1 − estimation accuracy) as the loss function would be the optimal binning of context [11], but building it in our case is computationally prohibitive[2]. The second method, which we use, performs clustering on the outcome distances to determine the bins. We use k-mean clustering based on the 2-norm distance of the normalized Laplace-corrected usage vectors to create the bins. Each usage vector is in the form of $\{P(g_1|x), P(g_2|x), …, P(g_{100}|x)\}$. One must note that in order to preserve the integrity of results when binning, it is necessary to separate the training data used for creating the bins from the testing data.

## 4. Context Dependency of Phone Usage

In this section, using the context and usage traces, and our formal definition of context dependency, we present a series of interesting findings regarding the context dependency of web, phone, and app usage. As presented in Section 3, we utilize prediction accuracy as the metric for evaluating context dependency. We note that context-based prediction of usage is extremely challenging due to the distribution usage that closely resemble the power law. We can see in Figure 2 that the most popular usage of each user constitutes a major proportion of their usage, and much higher than the next most popular usage, and so forth. Consequently, even given context based evidence, it is improbable for the posterior probability of any usage to rise above the more common usage.

As mentioned in Section 2, applications and services in which the cost of a false negative is considerably higher than false positive can benefit from multiple responses in the form of $\hat{g} = \widehat{g_1} \cup \widehat{g_2} \cup ...$ . For example, for an app preloading application, the system might preload multiple apps to reduce the chance of a miss. Accordingly, we consider the case for multiple responses as well as the single response case.

### 4.1 Individual Context

We have studied the performance of individual context by measuring the estimation accuracy versus number of context bins, as shown in Figure 3. One context bin would mean no context, i.e. always returning the most likely result(s). We utilize leave-out-one cross-validation (LOOCV) to preserve the integrity of our results. LOOCV removes the test sample from the training data set used to calculate posterior probabilities. We utilize Simple Binning in order to keep the meaning of bins easy to understand. Recall from Section 3.4 that for categorical context, Simple Binning utilizes the top *n-1* categories and an 'other' category. For continuous context, it simply discretizes it into *n* bins.

In this section, we present our findings regarding the performance of each context, as well as the effect of the number of bins. Note that different context have a widely diverse range of effectiveness, depending on usage and the number of acceptable responses. Also, we see that more bins initially improves performance, but after a point will hurt performance. The reason is that even though more bins can allow finer molding of the model, hence a more accurate calculation of $P(g|x)$, it would result in fewer samples per bin, increasing noise and reducing accuracy.

An important finding not inherently obvious in the figures is that an increase in the number of context bins is useful only as long as there are a reasonable number of samples to reliably calculate the posterior probability of each bin. As a rule of thumb, there should be more than ten samples per bin, even though we are mitigating the ill effects of too few samples using Laplace correction. As shown in Table 1, there are on average 700 website visits per user. Therefore, it is unsurprising that in particular for equal cluster size contexts, there are diminishing results in going over 10-50 context bins. On the other hand, since there are over two thousand phone call samples, increasing the number of context bins is fruitful up to 100-200 bins, where the returns are diminished. This shows that the number of context bins should be not preset, but dynamically adjusted by the system to ensure a reasonable number of samples per category.

We next provide findings specific to each context type shown in Figure 3, using the better performing discretization method for each context, i.e. with and without the equal frequency constraint when applicable.

---

[2] One can, however, use heuristic methods such as CHi-squared Automatic Interaction Detector (CHAID) [32] to build a (suboptimal) decision tree.

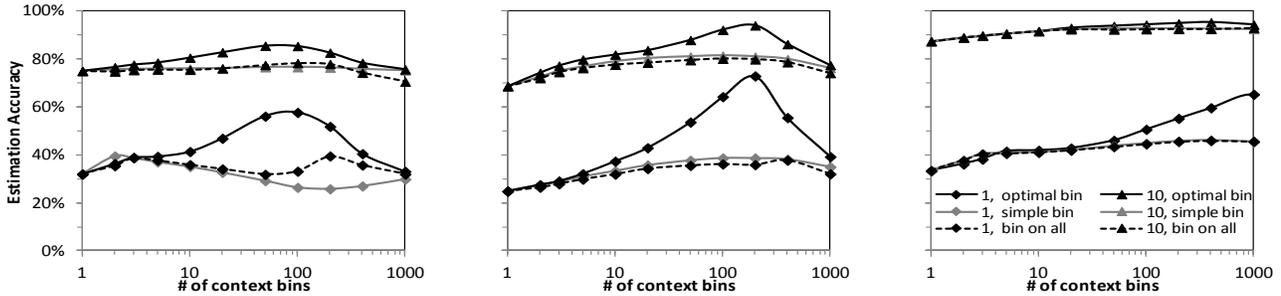

**Figure 5: Supervised Binning, performed individually for each user, can greatly increase the accuracy of context-based usage estimation. Estimation accuracy of web (top), phone call (middle), and application usage (bottom), calculated using the Bayes method, for one, three, and ten acceptable responses. Compared to Supervised Binning on all users' data, and individualized simple binning. One bin means no context information was used.**

**Time&day:** Recall that we found separating weekends from weekdays increases performance, compared to treating all days as the same. We also found that equal frequency discretization of time&day performs better than equal width discretization. The effectiveness of time&day levels off early, when the number of bins is extended beyond ~20.

**Movement:** Similar to time&day, equal frequency discretization of accelerometer power performs slightly better than equal width discretization. Interestingly, a relatively high number of bins (e.g. 100) are most effective here. This is in contrast with our original expectations that a small number of bins, e.g. to account for moving and non-moving states, would be sufficient. This finding suggests that accelerometer power can and should be used *as a signature to classify a user's detailed state*, and not merely as an indicator for whether they are moving or not. Previous research has shown a similar phenomenon with ambient sound for the purpose of room level localization, i.e. SoundSense [33].

**GPS Location:** In contrast to the single dimension contexts, location performed best without the equal frequency cluster constraint. We believe this is due to the equal frequency constraint artificially breaking down meaningful clusters in order to satisfy the sample size constraint. Interestingly, without the equal frequency constraint, a larger number of bins do not reduce performance. We believe this is due to the extra clusters mostly absorbing outliers, instead of breaking down meaningful clusters.

**Cell ID Location:** As each cellular cell spans a large coverage area, most of our users' lives were under a small number of cell IDs. Therefore, there is little to gain from increased number of context bins. Note that as cell ID is already discrete, discretization doesn't apply here.

**Prior usage:** We can see that all three types of usage are most dependent on the prior usage of the same type, versus other forms of prior usage. For example, phone calls are more dependent on the prior phone call. Yet, we see other types of usage are also more or less good indicators. The only exception was that prior app usage seems to have no effect on web usage.

### 4.2 Combinations of Context

By observing the performance of the three types of usage, we can see that combination methods are very useful when a number of meaningful context sources are present. The max-rule consistently outperforms the mean-rule for our data. The mean-rule is known to perform well when classifiers are noisy and highly dependent. Therefore, we conclude that that different context sources are not noisy or highly dependent. The surprising fact that Bayes often outperforms other combinations is another indicator that our context sources are either not highly (conditionally) dependent, and/or their dependencies are distributed evenly [29]. On the other hand, even though we use Laplace Correction to reduce the impact of data sparseness, the performance of Bayes is reduced when there are more bins (with fewer samples and therefore more noise). Indeed, it is well known that Bayes is highly susceptible to noise.

We note that as expected, and as confirmed by the traces (not shown), treating multiple context sources in a multi-dimensional manner results in virtually no improvement in estimation accuracy. This is unsurprising, as even with only 10 context bins, there were less than 1% of samples in any given context. More importantly, most samples belonged to bins that each contained less than 0.1% of samples.

### 4.3 Seasonal Variation

Our long-term traces allow us to answer an important hypothesis regarding context dependency: how significant is the effect of the user's seasonal variation on estimation accuracy? In other words, whether shorter durations may show higher context dependency due to the significant seasonal changes in user behavior [19], or would fewer samples in shorter durations reduce estimation accuracy.

We find that the context-dependency of usage remains relatively constant even for durations of one to three months, as shown in Figure 4. Furthermore, the fact that the performance is not significantly reduced for shorter durations indicates that a smaller data set, e.g. a month or more, would be sufficient for context-awareness. To this end, we present the same analysis as the prior section for the Bayes combination method, but instead of only calculating it for the entire 12 month duration of the study, we calculate it over one, three, and six month durations as well. For each duration and usage, we select the best number of bins, and present the average estimation accuracy of the multiple partitions resulting from each duration, e.g. the results from the four 3-month durations are averaged together.

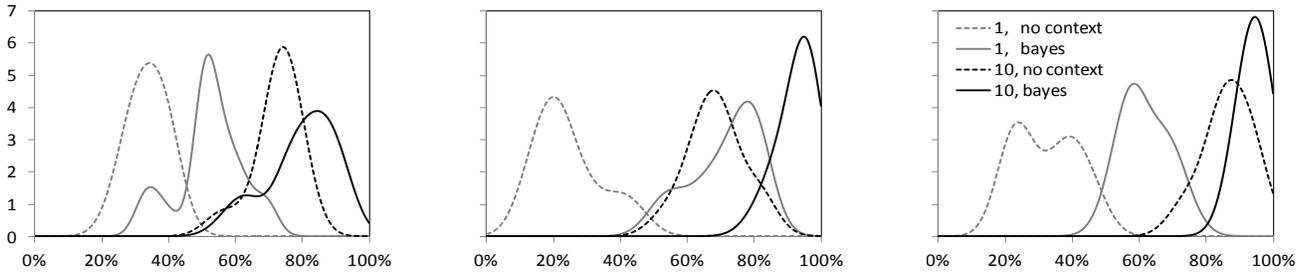

**Figure 6:** User diversity in context dependency is shown by the Kernel Density Estimation (KDE) distribution of estimation accuracy for one and ten responses. Web (left), phone (middle) and application usage (right).

### 4.4 Supervised Binning

As mentioned previously, there is an inherent tradeoff in choosing the number of context bins; more bins can allow finer molding of the model, and more accurate results, but it would at the same time result in fewer samples per category, increasing noise and reducing accuracy. Supervised Binning has the potential to make the best of both worlds, by identifying and binning together categories that have similar outcomes. We have studied the efficacy of binning, and have shown that it can greatly increase the accuracy of context based usage estimation.

We apply Supervised Binning as follows. For continuous context, we first discretize it into ten times the categories, to allow for sufficient freedom for the binning algorithm while avoiding overfitting. It is computationally prohibitive to apply LOOCV for binning, as it would require recalculation of the binning for each test case. Therefore, we utilize two-fold cross validation, splitting the data into two six-month durations, and use the first six months for training and the second for testing, followed by the opposite.

We present significant performance increase of Supervised Binning in Figure 5. We use the Bayes combinatory method, since it produced the best results (Section 4.2). For fair comparison, we also show the performance of Simple Binning calculated using the same two-fold cross validation.

One interesting question is whether it is necessary to perform supervised binning individually per user, or are there inherent features in context that are common between users, and can allow the supervised binning to be performed once for all users, i.e. using data from all users. Our results confirm the former; even for our small, relatively homogeneous population, supervised binning using data from all of our users fails to improve accuracy over simple binning.

### 4.5 User Diversity

The long-term traces allow us to analyze the diversity in the context dependency of different users. i.e. whether some users have more diverse usage, and whether some users' usage is more context dependent. We utilize the Kernel Density Estimation (KDE) to present the distribution of estimation accuracy among our participants, and compare it to the case without context information (only one bin), for one and ten acceptable responses, as shown in Figure 6. The estimation accuracy is calculated using our best methods, i.e. Bayesian combination and Supervised binning, and we empirically set the KDE bandwidth to 0.05.

Figure 6 shows that while there is considerable diversity among participants' usage patterns, all of them show context-dependency in their usage. Furthermore, the top one and ten usage cases, which serve as the baselines for context dependency, constitute a significant share of usage for all participants. Finally, we note that among the three principal usage we studied, the estimation accuracy for web usage had a much higher diversity of context dependency, compared to its non context case. This shows that there is more diversity in the context dependency of web usage, compared to phone and app usage.

### 4.6 Prior Usage Context

The methodological classifier combination approach to the data sparseness challenge of context awareness, presented in Section 3.2.2, allows the use of prior usage context in addition to sensor context. Recall from Section 2.3.2 that usage context refers to the last used websites, phone calls, and apps, and sensor context refers to the measurements of the device's sensors, i.e. time&day, cell ID, motion, and GPS. In this subsection, we examine the effectiveness of prior usage context of varying depth for estimation accuracy. The depth of prior usage context is defined as how many prior usages are considered and combined along with the sensor context. A zero depth means no usage context is utilized. For the depth of one, the last used prior usage for web, phone, and apps is used, similar to other sensor context. So far, we have only utilized prior usage with a depth of one. Higher depth usage context can be treated in two ways. First, n-th depth usage context can be simply presented as yet another single dimension usage context, which will be handled and combined similar to other context, as described in Section 3. Second, is to treat each n-depth usage context as an n-dimensional vector. Compared to the first approach, this method suffers from sparseness, due to the curse of dimensionality.

We have found that estimation accuracy drops when the depth of historical context is increased beyond one. This limitation stems from inherent limitations of classifier combination for dependent data. Figure 7 shows the best estimation accuracy achieved through Bayesian combination and Supervised Binning, for different depths of prior usage context for phone usage. Web and app usage were similar, and are therefore not shown. It shows that after the depth of one, the gains start to diminish. This is due to the limitations of classifier combination; the additional information provided by depths higher than one cannot offset the error induced by their dependence Also, for the multi-dimensional treatment of n-depth context, due to its sparseness for most of the samples, Laplace Correction simply returns the average a posteriori probability as the conditional a posteriori probability. In turn, this adds dependency and reduces performance.

**Table 2. Ordering & measured energy cost of context**

| Type of context | Order | Energy cost |
|---|---|---|
| Prior Usage, time&day | 0 | negligible |
| Accelerometer | 1 | 1.65 J |
| Cell ID | 2 | 1.2 J |
| GPS location | 3 | 50 – 300 J |

In order to better support historical usage context, for depth of greater than one, it is obvious that the dependency challenge needs to be addressed. Therefore, we propose and evaluate the following method for incorporating historical context: Treat usage context with a depth higher than one as multidimensional, but for each classification event and each type of usage, choose the highest depth that has more than $m$ training samples. This method is referred to as auto-depth. The constant $m$ is the minimum number of training samples deemed sufficient for an accurate a posteriori probability calculation. We set $m$ to 10, based on the findings in Section 4.1. We find that the automatic depth selection approach works well, in terms of avoiding a performance drop, i.e. more information doesn't hurt. However, there is no measurably significant performance increase after the depth of one for the measured data.

## 5. Cost-Aware Context Combination

Our findings so far attest to the performance and usefulness of context-based usage estimation. However, obtaining and processing context information can incur significant energy costs. Many ad-hoc methods, sometimes themselves context based, have been employed to reduce energy cost of context awareness while satisfying the system designers cost / accuracy tradeoff. These methods typically reduce the frequency of accessing energy hungry context sources, or to avoid using them altogether, substituting them by lower cost context sources.

In this section, we address this challenge through a methodological framework, SmartContext. It takes advantage of the methodology presented in Section 4, and builds upon the general problem of budgeted observation selection in the operations research community, to automatically optimize the energy cost of context-based estimation, while satisfying the accuracy requirements and tradeoffs that the designer sets for each estimation event. We focus on energy costs, but other definitions of cost may be used as well.

### 5.1 Assumptions and Requirements

SmartContext takes advantage of and is compatible with any classifier combination method, as long as the combination method can provide the MAP estimate and estimation accuracy using any combination of context sources with small processing overhead. All three classifier combination methods we evaluated have these features, but we will use Bayesian as it performed best.

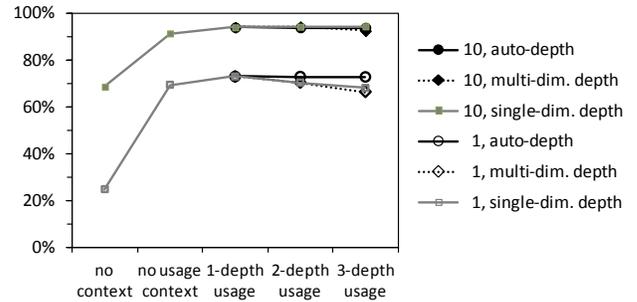

**Figure 7:** Automatic depth selection is necessary to efficiently utilize prior usage context with depth greater than one. Effectiveness of prior usage context in increasing estimation accuracy, for phone usage, vs. depth of prior usage context for one and ten acceptable responses, averaged among the 24 participants.

SmartContext selects context sources in order to meet an estimation accuracy, set by context-aware applications and services, for every estimation event. SmartContext requires that the training data set be available for all context sources, i.e. $P(g|x_n)$ for all $n$. SmartContext requires the cost, or the expected cost, of utilizing each context source to be known in advance. Note that the costs can be independent or dependent on each other. Further, the costs of some context sources are negligible. Therefore, they will be always utilized, limiting selection to context sources with non-negligible costs.

### 5.2 Operation

SmartContext's operation consists of two main steps. The first is determining the ranking of context sources. In order to keep processing costs in check, this ranking must be pre-calculated, but can be always static, or can be dependent on the context information gained at any step. In the next section we show that a static solution is both practical and performs well. In this case, the ranking needs to be performed only once. The second step is the energy aware combination of context. This has negligible overhead, and is performed dynamically for every estimation event according to the requirements and tradeoffs of the context-aware application or service.

Once the ranking is determined, the energy aware combination of context works as follows. For each classification event, SmartContext starts combining multiple sources of context information one by one, in the ordering determined in step one. Note that this can be done with minimum processing overhead, and for any combination of context sources, as explained in Section 3.2.2. After running the classifier combination with each additional context source, it checks the criteria of the requesting application or service, for that estimation event. In the evaluation presented here, a fixed minimum estimation accuracy for every esti-mation event is utilized. However, the application or service may set a different accuracy requirement for each estimation event, or even consider the expected cost of accessing the next context source, in determining when to settle with the current estimation accuracy and stop accessing more context sources. SmartContext assures the target estimation accuracy for each estimation event, as long as it is possible to reach that accuracy, while spending no excess cost in acquiring unnecessary context. In other words, in some conditions, no additional costly context is used, while in more uncertain conditions, SmartContext may use up to all the available context sources. The pseudo-code description of SmartContext is shown in Figure 10.

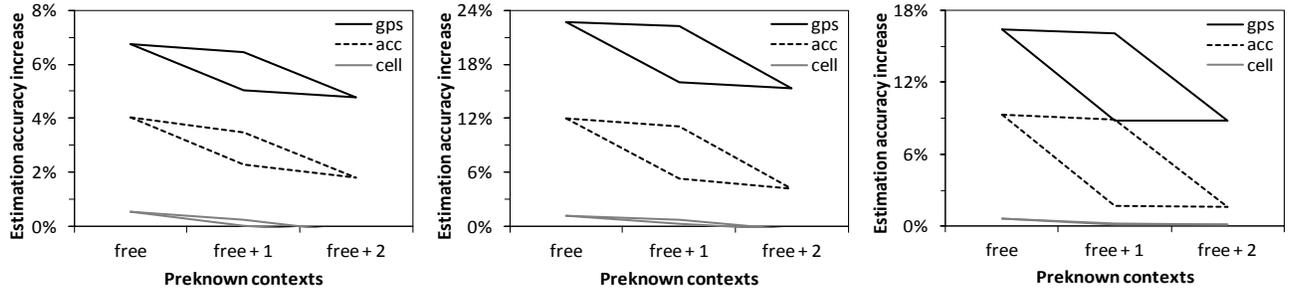

**Figure 8: Submodularity of estimation accuracy; the average estimation accuracy gain resulting from combining a certain context decreases if more contexts were known (combined) beforehand. Free indicates time&day and previous usage. Free + 1 and free + 2 indicate one or both of the remaining two contexts (except the one to be added). Left: web usage. Middle: phone call usage. Right: app usage**

### 5.3 Ranking of Context Sources

The ranking of context sources is analogous to a well studied problem in artificial intelligence and operations research, which can be defined as follows:

*How to select a subset, X, of possible observations (i.e. predictors or information sources) V, that most effectively reduces uncertainty and maximizes information gain?*

#### 5.3.1 Review of existing methods

Solutions toward this challenge are based on submodularity, an important property of the information gain from multiple observations [34]. Submodularity is also intuitively named as the diminishing returns property. It states that the information gain from an observation helps more if one has a smaller set of observations so far. Vice versa, the information gain from an observation helps less if it is added to a larger set. This can be formally presented as follows. The set function $F : V \rightarrow \mathbb{R}$ is submodular if

$$F(A \cup X) - F(A) \geq F(A' \cup X) - F(A')$$

for all $A \subset A' \subseteq V$, $X \notin A$, i.e. adding $X$ to a smaller set, $A$, helps more than adding it to a larger set, $A'$. The general problem of maximizing submodular functions is NP-hard [35], and general algorithms are unable to provide guarantees in terms of processing time [36], unless there are certain assumptions, e.g. selecting a subset among a fixed tree ordering of possible observations [37]. However, artificially imposing such a dependency tree is heuristic in nature and can reduce performance, e.g. in [38]. Further, calculating the maximum-likelihood dependence tree requires assumes pre-measured mutual information between unit cost observations are available [39], neither of which are applicable to our case.

Therefore, it is common practice to use the greedy (myopic) solution towards this selection problem [40]. The submodularity property ensures that such greedy solutions are near-optimal, typically with provable constant factor performance guarantees. The greedy solution, assuming a unit cost for all observations, selects the observation with the most information at every step, i.e. the marginal increase $F(A \cup X) - F(A)$ is maximized. For this case, in [34], Nemhauser et al. prove that any set of equal-cost observations selected in this manner performs, at worst, a factor of $(1 - 1/e)$, compared to the optimal set. More recently, the operations research community has proved the same bound when observations have different costs. Krause et al. prove that for independent costs, the greedy solution selects the observation with the maximum *cost-effectiveness* at every step, i.e. the marginal increase divided by cost of observation, $(F(A \cup X) - F(A))/cost(X)$, is maximized [40]. Furthermore, the same work proves that approximation algorithms are unable to provide guarantees better than a constant factor of $(1-1/e)$, i.e. $(k \cdot (1-1/e))$. We therefore base our work on the solution provided by Krause et al. [40].

#### 5.3.2 Ranking Mobile Context

SmartContext is based upon the greedy method described in the earlier section, guaranteeing [40] a performance bound of $(1 - 1/e)$. However, the performance guarantee requires two assumptions. First, the costs of context sources (observations) must be independent from each other. Indeed, mobile context sources typically have independent energy costs, as was our case. Second, the submodularity or diminishing returns property must hold for our utility function (estimation accuracy). While this appears a reasonable assumption, it is necessary to verify it.

Existing work typically use entropy as their utility function, and either assume that it is submodular [34], or prove that it is submodular under an assumption of independence [40]. We experimentally verify the overall submodularity of estimation accuracy. For this purpose, it is necessary to show that the estimation accuracy gain resulting from adding (combining) any context source decreases if more context sources were known (combined) beforehand. Note that

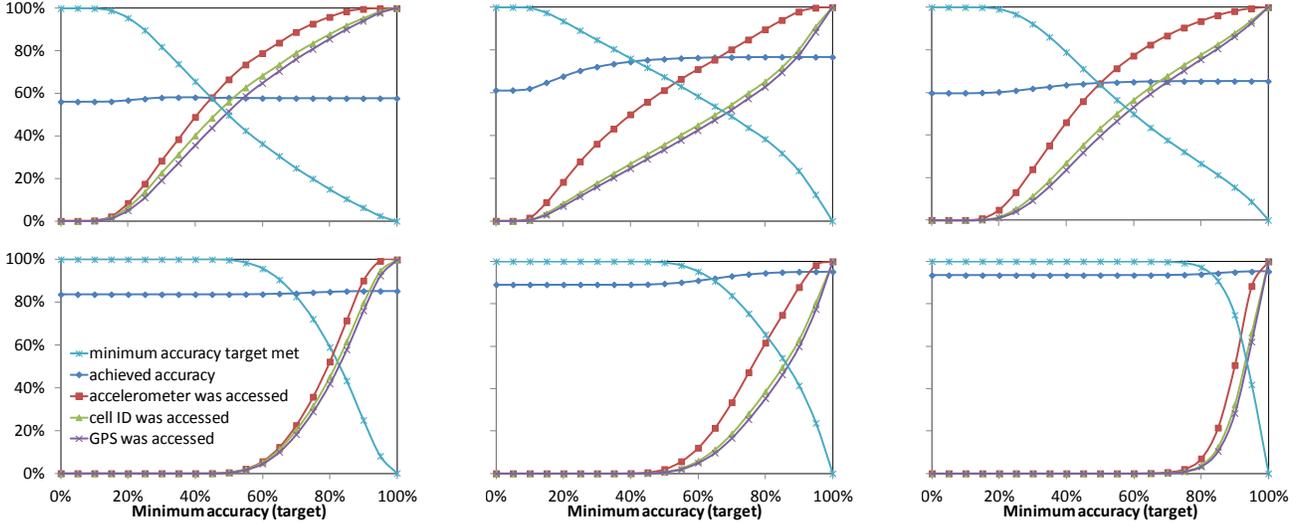

**Figure 9: SmartContext's performance for web (left), phone (middle), and application (right) usage, for 1 (top), and 10 (bottom) responses. We show, for a range of minimum accuracy targets, how often costly context is accessed, how often the minimum accuracy target is met, and the overall average estimation accuracy.**

```
DetermineCostPerformanceOrdering(context_sensors)
ForEach (sensor) in (sorted_free_context) do {
    accuracy, usage = CombineNextContext(sensor)
}
ForEach (sensor) in (sorted_costly_context) do {
    If AppConditionMet(accuracy, usage, costs[])
        Exit Loop
    accuracy, usage = CombineNextContext(sensor)
}
Return (accuracy, usage)
```
**Figure 10: Pseudo-code for SmartContext**

since SmartContext assumes that free context is always utilized, it is necessary to verify the submodularity only among costly context. Figure 8 shows the estimation accuracy gain for Cell ID, acceleration, and GPS location. Therefore, we conclude that the greedy approach works well for context awareness. In this case, the best ordering is obtainable by ranking the context sources according to their cost effectiveness. In this case, the cost effectiveness of each context source is the marginal estimation accuracy increase divided by its expected energy cost, i.e., $(F(A \cup X) - F(A))/cost(X)$. The expected energy cost can be pre-measured by the system designer, as in our case, or can be measured automatically in software as in [41]. The energy costs of acquiring context on the iPhone 3GS are presented in Table 2. This desertion measured the estimation accuracy of each context source individually in Figure 8. The resulting ranking is shown in Table 2. Note that due to the often significant difference in the energy cost of context sources on mobile devices, their ranking becomes close to the order of their energy cost.

Finally, we note that due to the relatively limited number of costly context sources on mobile devices, it is also possible to simply perform a thorough search, calculating the performance of SmartContext under all possible orderings of context sources. For our case, there are three costly context sources, resulting in a total of *3! = 6* possible rankings. Unsurprisingly, the rankings we obtained using the through search are in fact, the same as the greedy ranking calculated in this section, for each of the experimental cases of Section 5.4.

### 5.4 Evaluation

We have evaluated SmartContext using the in-situ traces. Figure 9 shows, for different (minimum) target accuracies, how often each context source is utilized for Cell ID, Accelerometer, and GPS, and how often the target accuracy is achieved, as well as the overall average achieved estimation accuracy. Note that SmartContext always uses time&day and previous usages, as we assume they are available for free. We can see that even without energy hungry context sources, and only using free context, it is often possible to achieve good estimation accuracy, and the additional accuracy provided by the costly sources are incremental. This is expected, because each source of context has a small incremental value, as shown in Section 4.1, and because submodularity ensures diminishing returns, as shown in Section 5.3 and Figure 8.

Yet, we show significant energy savings are possible, with very little sacrifice of accuracy. . For example, for web, phone, and app usage respectively, for one acceptable response, setting the (minimum) target accuracy to 25%, 50%, and 50%, achieves 89%, 67%, and 61% energy saving, while providing overall estimation accuracy within 1% of the case using all context sources. For ten acceptable responses and 75%, 80%, and 85% (minimum) target accuracies, respectively, the energy savings are 71%, 65%, and 89%, while again achieving overall estimation accuracy within 1% of the case using all context sources.

Note that as the ordering of context and the posterior probabilities are pre-calculated, e.g. during charging, they do not add to the overhead of SmartContext during operation. Further, the combination algorithms require little processing, therefore SmartContext has a negligible overhead.

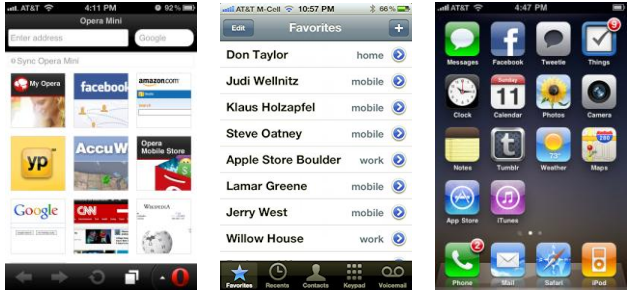

**Figure 11: Sample applications that can benefit from a dynamic context-aware selection vs. currently static selections. Left: Browser displays a list of bookmarks when launched. Center: Favorite phone contacts. Right: Apps on the home screen, and the always-visible quick launch bar (bottom row).**

# 6. Sample Applications

The context dependency of mobile usage not only provides insight to the social behavior of humans, but can be utilized in many applications, such as those in Figure 11. In this section, we provide a brief description of several such applications, and their expected performance gains based on our best performing methods, i.e. using Supervised Binning and Bayesian combination.

*Web bookmarks*

It is known that a few websites account for most of the typical user's usage [42]. Accordingly, some browsers, e.g. Opera, offer a list of favorites or home screen that is configurable by the user or automatically generated. This would provide with access to the user's most common websites. Others provide a user configurable home page that is automatically loaded when the browser is run. Similar to Section 4, we focus on individual domains, instead of pages within a domain.

A context-aware web favorites list is a sample application that can present more likely choices to the user according to their context. Our findings, presented in Figure 12, show that a context-based solution for providing the user with either a single default home page, or a list of 10 websites, significantly outperforms an ideal static selection, with a miss rate of 15% vs. 25% for a list of 10 websites, and 42% vs. 68% for the single home page, and less than half the ideal static solution. Interestingly, the ideal static list of 10 favorite websites outperforms the 10 most recently visited.

*Phone favorites list*

In order to assist users in making phone calls, phones typically provide the user with a redial button, a list of recent phone calls, and/or a user configurable favorites list. For example, the iPhone used in our study provided a list of recent phone calls as well as a user configurable favorite contacts list. A context-aware phone favorites list is a sample application that can present more likely choices to the user according to their context.

On average, a static list of each user's top 10 contacts has a miss rate of 32%, and a recent call list has a miss rate of 28%. On the other hand, a context aware favorites list can reduce the users' need to go through their phonebook by approximately five fold, to 6%. Furthermore, the miss rate of a redial button is 64%, but the context-based dial button has a miss rate of 27% (Figure 12).

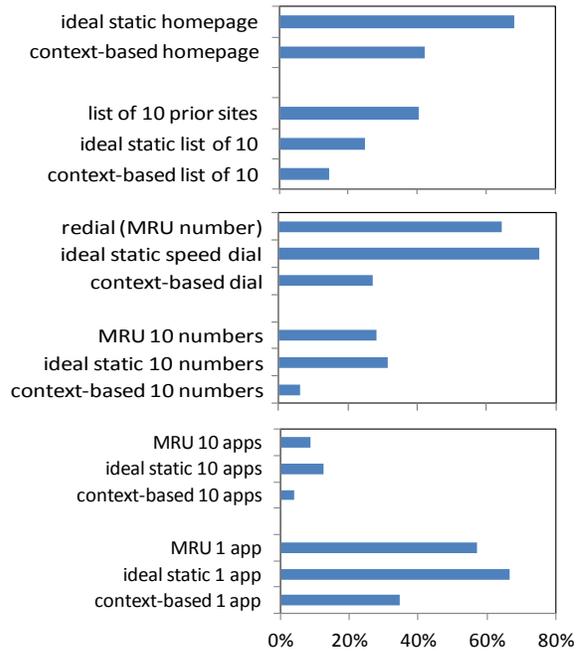

**Figure 12: Performance of context-based applications, presented as miss rates. Top: web bookmarks. Middle: phone favorites. Bottom: application launching or preloading**

*App Quicklaunch and Preloading*

Most phones often have a list of apps that are more readily available for users to run, i.e. quicklaunch. The iPhone provides room for four such apps, which are readily available on any page of the home screen, and users can also organize their apps so the most common are placed in the first page. A context-aware quicklaunch list is a sample application that dynamically updates the quicklaunch list according to the users' context. Our findings show that it would have a miss rate of 16%, compared to the 39% miss rate of the ideal static quicklaunch, an improvement of three times. For 10 apps, the miss rate is just 4%, compared to 13% for the static case (Figure 12).

Preloading is another possible application, where context-based estimation of the application to be used can enhance performance. App preloading, including context-based methods have been widely studied in the past [43]. We have measured the app launch times on the substantially faster iPhone 4. The measurements were repeated three times for each app, and we excluded content load times, if applicable. For the without preloading case, each app's process was manually terminated between the runs. For the preloaded case, we started the app once, and closed it before the measurement run. Without preloading, the average load time was 2.0 seconds (median = 2.1, deviation = 0.5). With preloading, the load times were 0.6 seconds for all the measured apps. These measurements show that, on average, preloading can improve app load times over three fold. We note that the iPhone and many other platforms utilize a most recently used algorithm to keep multiple apps in memory, given memory limitations. We compare our performance to the MRU algorithm, and show that the miss rate for 10 preloaded apps, the improvement is from 9% to 4% (Figure 12). Furthermore, for 3 preloaded apps, the miss rate is reduced from 31% to 17%, almost half.

# 7. Related Work

Prior work (e.g. [2]) also define context dependency as a set of strict or probabilistic rules and relations between context(s) and the outcome. Others design and implement frameworks for sensing and processing context information [44, 45]. For a survey, see Baldauf et al. [46]. These work attest to the significance and usefulness of context.

Context information, has in the past been widely used for specific applications such as implicit user interaction (e.g. by Schmidt [5]) and information delivery (e.g. a reminder system by in [6], a tourist guide in [7], and content adaptation in [3] and [4]). For a survey of such cases, see Chen and Kotz [47]. Others have presented system mechanisms based on context information, e.g. estimating and predicting wireless network conditions [10], network routing [48], battery management [8, 49], and energy efficient GPS duty cycling [12, 13]. Further, Eagle and Pentland have shown that device usage patterns are indeed structured and predictable [50]. These designs and others depend on the context dependency of device usage, and show significant, quantified, performance gains by exploiting context.

A number of other work depend on knowledge regarding usage to perform. For example, the work in [10] predicts network conditions to choose the best network interface, but assume network usage is pre-known, even though it depends on the behavior of applications, services, and the user. As another example, the authors of [8] show that battery consumption is context dependent. The authors of [9] further research this problem by focusing on phone call usage, and show that call lengths, and therefore their energy consumption, are context dependent. Further, Eagle and Pentland have shown that device usage patterns are indeed structured and predictable [50]. The usefulness of context has been so significant that many researchers have designed and implemented frameworks for the specific task of sensing and processing context [44-46].

Our work presents a methodological solution for using multiple and various sources of context while managing their energy costs, and presents a formal definition of context dependency as well as practical methods to calculate it. We abstain from focusing on a single application or service, yet we provide practical insight into the relationship between context-dependency and the performance achievements of individual applications.

A number of recent research have dealt with reducing the cost of acquiring context. These work attest to the challenge of energy efficiency in context awareness, but typically focus on single applications and/or static configurations. They use one or more of the following three techniques to reduce energy cost, while retaining acceptable performance. First: *frequency reduction*, as in [12-14] reduces the sampling frequency of energy hungry context sensors. Second: *sensor substitution* utilizes lower energy cost context instead of energy hungry ones, as in [10, 12, 13]. Third: *sensor elimination* attempts to utilize a subset of sensors. We take the third approach in SmartContext, but unlike previous work that focus on and take advantage of the properties of a specific application, such as activity detection [15-18], we provide a generic framework for system designers to dynamically or statically optimize the cost / accuracy tradeoffs of context awareness. A more general problem of selecting the most effective subset of sensors, also referred to as observations or predictors, has been the focus of decades of research in the artificial intelligence and operations research communities. These work focus on the optimization of information, typically defined as either joint entropy or information gain (delta entropy). It is common practice to use the greedy (myopic) solution towards this selection problem [40, 51, 52], with guaranteed performance bounds, due to the processing complexities of finding the optimal solution [35-37]. SmartContext builds upon the greedy solution of Krause et al. [40], but is adapted to using estimation accuracy instead of information gain.

There has been several research utilizing phone logging, e.g. [53-56]. Compared to our traces, they collect very limited context information due to privacy concerns and battery lifetime limitations. We have overcome these challenges by the careful design and implementation of the study, and have collected unprecedented data.

Finally, there has also been considerable research on determining user state from context information e.g. physical activity [57]. In this work, we abstain from extracting user state, either directly or as an interim stage, and instead focus on the relationship between device usage and context information.

# 8. Summary and Conclusion

We have found that estimation accuracy based on MAP estimation is a practical application agnostic measure for context dependency, yet can provide insight into the real-life performance of many applications, several of which are briefly presented here. These applications attest to the effectiveness of context for estimating usage, and highlight the practical value of estimation accuracy as the measure of choice for context dependency.

We have also found that due to the power law distribution of usage, estimating mobile usage is very challenging. Yet, we show that careful yet methodological treatment of multiple sources of context, e.g. combination, discretization, binning, can greatly increase estimation accuracy. In particular, we have found that 1) it is necessary to maintain a reasonable number of usage samples in each category, i.e.

no less than ten, and equal frequency discretization of single dimension context helps achieve this. 2) Classifier combination methods can address the data sparseness challenge when utilizing multiple context sources, and Bayesian combination works best, even though the contexts are dependent. 3) Individualized supervised binning greatly improves estimation accuracy by keeping a more samples in each bin while allowing the fine molding of bins.

Finally, even though the energy cost of some context sources can be a substantial challenge for context based applications. We address this challenge through the SmartContext framework, which ensures using only as much context sources to meet a minimum accuracy set by the application designer for each estimation event. We show that SmartContext can achieve an estimation accuracy within 1% of the maximum possible accuracy, while reducing energy costs by 60% or more.